\documentclass[iop]{emulateapj-rtx4}
\usepackage{natbib}
\usepackage{graphicx}                    
\usepackage{amssymb,amsmath}
\usepackage[usenames]{color}

\shorttitle{Plasma composition in an active region}
\shortauthors{Baker et al.}

\begin{document}

\title{Plasma composition in a sigmoidal anemone active region} 

\author{D. Baker\altaffilmark{1}, D.~H. Brooks\altaffilmark{2}, P.~ D\'emoulin\altaffilmark{3}, L. van Driel-Gesztelyi\altaffilmark{1,3,4},   L.~M. Green \altaffilmark{1}, K. Steed \altaffilmark{5}, J. Carlyle\altaffilmark{1}}
\altaffiltext{1}{University College London, Mullard Space Science Laboratory, Holmbury St Mary, Dorking, Surrey, RH5 6NT, UK}
\altaffiltext{2}{College of Science, George Mason University, 4400 University Drive, Fairfax, VA 22030, U.S.A.}
\altaffiltext{3}{Observatoire de Paris, LESIA, CNRS, UPMC Univ. Paris 06, Univ. Paris-Diderot, Meudon, France}
\altaffiltext{4}{Konkoly Observatory, Hungarian Academy of Sciences, Budapest, Hungary}
\altaffiltext{5}{Centre for mathematical Plasma Astrophysics, KU Leuven, Celestijnenlaan 200B, 3001 Leuven, Belgium}

\begin{abstract}

Using spectra obtained by the EIS instrument onboard \emph{Hinode}, we present a detailed spatially resolved abundance map of an active region (AR) -- coronal hole (CH) complex that covers an area of 359$\arcsec$ $\times$ 485$\arcsec$.  The abundance map provides first ionization potential (FIP) bias levels in various coronal structures within the large EIS field of view.   Overall, FIP bias in the small, relatively young AR is 2--3.  This modest FIP bias is a consequence of the AR age, its weak heating, and its partial reconnection with the surrounding CH.  Plasma with a coronal composition is concentrated at AR loop footpoints, close to where fractionation is believed to take place in the chromosphere.  In the AR, we found a moderate positive correlation of FIP bias with nonthermal velocity and magnetic flux density, both of which are also strongest at the AR loop footpoints.  Pathways of slightly enhanced FIP bias are traced along some of the loops connecting opposite polarities within the AR.  We interpret the traces of enhanced FIP bias along these loops to be the beginning of fractionated plasma mixing in the loops.  Low FIP bias in a sigmoidal channel above the AR's main polarity inversion line where ongoing flux cancellation is taking place, provides new evidence of a bald patch magnetic topology of a sigmoid/flux rope configuration. 
\end{abstract}


\section{Introduction}

Elemental abundances of the Sun and their spatial and temporal variations are crucial to our understanding of the physical processes inherent to space weather as they can be used to probe the source regions of the solar wind (SW) and to trace the SW throughout interplanetary space.  Furthermore, composition can provide clues to the magnetic topology of the SW source regions as plasma on open magnetic fields can be distinguished from plasma confined in closed loops \citep{woo04,laming04,wang09}.  As demonstrated by \cite{lvdg12}, the magnetic topology of active regions (ARs) bordering coronal holes has significant implications for coronal outflows and the SW.

In a review of spectroscopic measurements of the abundances in the solar atmosphere, \cite{meyer85b,meyer85a} determined that the elemental abundance variation observed in the solar corona and SW compared with the photosphere strongly depends on the first ionization potential (FIP) of the element.  No other systematic trend could be identified with any other parameters, e.g.  mass and charge \citep{meyer91}.  Elements can be divided into low FIP and high FIP groups with the step at approximately 10 eV.  Those elements with low FIPs (e.g. Mg, Si, Ar, Fe) are enhanced in the corona and the SW by a factor of 3$-$4 over photospheric abundances whereas high FIP elements (e.g. C, O, Ne, S) maintain the elemental distribution of the photosphere.  This phenomenon is known as the FIP effect.   In order to characterize the abundance variations of the solar upper atmosphere, typically the corona, the FIP bias (FIP$_{Bias}$) is used to define the ratio of the elemental abundance in the solar atmosphere (A$_{SA}$) to the elemental abundance in the photosphere (A$_{Ph}$) such that FIP$_{Bias}$ = A$_{SA}$/A$_{Ph}$.

Though the composition of the photosphere is well-known and seemingly invariant, the composition of the corona and the SW varies substantially from structure to structure and with time.  In general, the elemental abundances of the corona are closely related to its morphology \citep{feldman02}.  The FIP bias of coronal holes (CHs) has been determined to be $\sim$1, i.e. photospheric in composition \citep{feldman93,doschek98,feldman98, brooks11}, whereas quiet Sun (QS) plasma was found to vary from $\sim$1.5$-$3.5, depending on the time in the solar cycle, the height above the limb, and the methods used to measure FIP bias \citep{feldman93,doschek98,warren99,landi06}.

The active Sun provides considerable FIP bias variation.  AR plasma confined to loops at the time of flux emergence showed photospheric composition \citep{sheeley95,sheeley96,widing97,widing01}, however, within a few hours, the FIP bias progressively increased to reach coronal abundances after two days \citep{widing01}.  The FIP bias levels of more established ARs ($\sim$ few days) were 4.8 to 5.9 \citep{widing95} and old ARs observed for at least seven days had FIP bias values ranging from 8 to 16 \citep{feldman92,widing95,young97,dwivedi99,widing01,feldman03,feldman04}.

External AR loops rooted in unipolar areas, known as spikes at the edges of ARs \citep{young97,young98} or Mg {\sc ix}  sprays \citep{sheeley96} in earlier studies, were measured to have very high values of FIP bias.  More recently, \cite{brooks11} derived a FIP bias of 3.4 for the upflow regions of AR 10978 over a 5-day period in December 2007.  The FIP bias derived by \cite{brooks11} of the upflows from the western side of AR 10978 was found to match the value measured in situ a few days later, thus providing observational evidence that upflows become outflows.  In general, measurements of the slow SW and solar energetic particles (SEPs) established that low FIP elements are enriched by a factor of more than 3$-$4 \citep{meyer85b,meyer85a,gloecker89,feldman92,reames98}, comparable to levels observed in ARs and streamers over the activity belt \citep{raymond97,feldman98}.  See the excellent review paper by \cite{feldman03} and the references therein for a more complete discussion of elemental abundances and their variations in the solar atmosphere.

\begin{deluxetable}{lcc}
\tabletypesize{\scriptsize}
\tablewidth{0pt}
\tablecaption{FIP Bias in solar structures}
\tablehead{
\colhead{Structure}            & \colhead{FIP Bias}    &
\colhead{Reference}}
\startdata
\bf{Coronal holes}             & $\sim$1, $\sim$2, 1.2 & 1, 2, 3, 4\\
\bf{Plumes}&9.4&5\\
\bf{Quiet Sun}                 & 1.5$-$2.0             & 1  \\ 
                               & $\sim$2, 2.3, 3.5     & 1, 6, 7\\ 
\bf{Streamers}                 & $\sim$3, $\sim$4      & 8, 2\\
\bf{Surges}                    & photospheric          & 9\\
\bf{Prominences/Filaments}&&\\   
Quiescent                      & 1.1$-$2.4             & 10\\
Eruptive	                   & photospheric          & 9\\
\bf{Flares}&&\\
Impulsive                      & $\sim$1               & 11, 12\\
B7.3$-$X1.5                    & 2.4$-$2.6             & 13\\
\bf{SEPs}                      & $\sim$4$-$6           & 14, 15, 16, 17\\
\bf{Active Regions}&&\\
Emerging                       & photospheric          & 12, 18, 19, 20\\
Established (a few days)       & 4.8$-$5.9             & 21\\
Old	 ($>$ 7 days)	           & 8$-$16	               & 17, 21$-$25\\
TR brightenings -- BPs$^{*}$   & photospheric          & 26\\
TR brightenings -- QSLs$^{**}$ & 2-4                   & 26\\
Core loop                      & photospheric          & 22\\
Spikes at edges/Mg {\sc ix}  sprays & 4-11 & 19, 22, 27\\
Upflows	                       & 3.4                   & 4\\
High speed blue-wing upflows&3$-$5&28\\
\bf{Solar wind}&&\\
Fast	                       & photospheric          & 29\\
Slow                           & coronal               & 29\\
\bf{Open B-field on disk}&	&\\
$>$ 14$\%$                     & 1.7$-$2.5             & 30\\
$<$ 7$\%$                      & 2.8$-$4.2             & 30\\
\enddata
\tablerefs{
(1) \cite{feldman93}; (2) \cite{feldman98}; (3) \cite{doschek98};
(4) \cite{brooks11}; (5) \cite{widing92}: (6) \cite{warren99};
(7) \cite{landi06}; (8) \cite{raymond97}; (9) \cite{widing86};
(10) \cite{spicer98}; (11) \cite{mckenzie92}; (12) \cite{widing97};
(13) \cite{sylwester13}; (14) \cite{meyer85b}; (15) \cite{meyer85a};
(16) \cite{reames98}; (17) \cite{feldman92}; (18) \cite{sheeley95};
(19) \cite{sheeley96}; (20) \cite{widing01};  (21) \cite{widing95};
(22) \cite{young97}; (23) \cite{dwivedi99}; (24) \cite{feldman03};
(25) \cite{feldman04}; (26) \cite{fletcher01}; (27) \cite{young98}; 
(28) \cite{brooks12}; 
(29) \cite{gloecker89}; (30) \cite{wang09}.  $^{*}$BP - bald patch separatrices.  $^{**}$QSLs - quasi-separatrix layers.}
\label{fip_bias}
\end{deluxetable}

Prior to the launch of the EUV Imaging Spectrometer (EIS; \citealt{culhane07}) onboard \emph{Hinode}, individual coronal structures were not easily distinguishable because of lower instrumental spatial resolution and limited fields-of-view (FOV; \citealt{feldman09}), consequently, previous elemental abundance studies provided FIP bias levels based on average compositions in coronal QS and CH regions or small patches of ARs \citep[e.g.][]{fletcher01}.  In this paper, we present a detailed spatially resolved abundance map of an AR--CH complex that covers an area of 359$\arcsec$ $\times$ 485$\arcsec$.   In the next section, we provide a brief description of the EIS observations of an AR--CH complex from 2007 October 17.  In Section \ref{fip_map}, we give an account of how the abundance map is derived using the S {\sc x} 264.223{\AA} and Si {\sc x}  258.375{\AA} lines observed by EIS.  Our results are presented in Section \ref{results} where we identify the FIP bias levels in the various coronal structures observed in the abundance map and we show correlation plots of plasma density, absolute value of magnetic flux density, non-thermal and Doppler velocities vs. FIP bias for the AR in the AR--CH complex.  We discuss the implications of our results in Section \ref{discussion} and draw conclusions in Section \ref{conclusions}.

\section{Overview of AR--CH Complex Observations }
\label{sec_obs}
\begin{figure*}
\epsscale{0.75}
\plotone{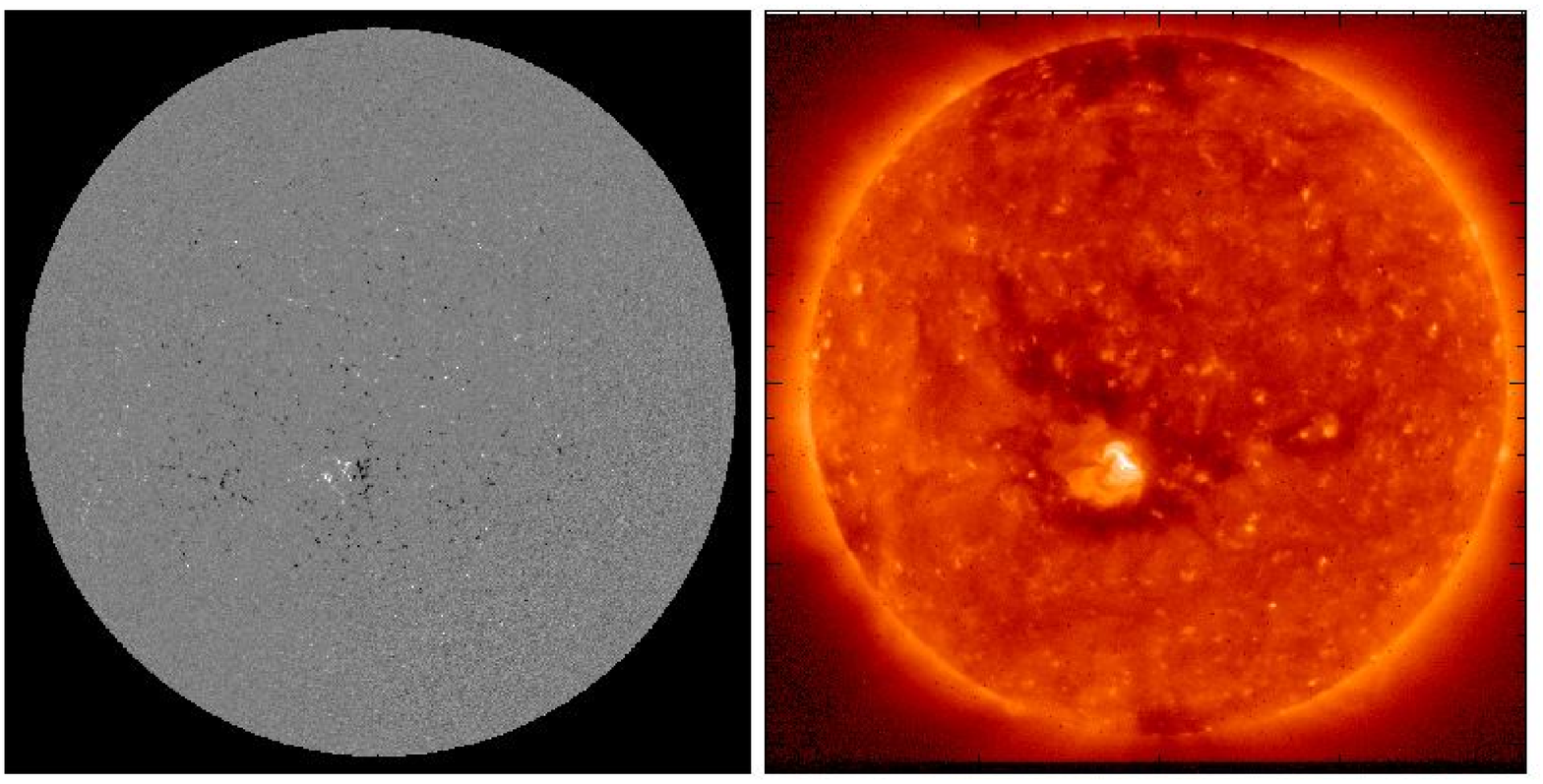}
\plotone{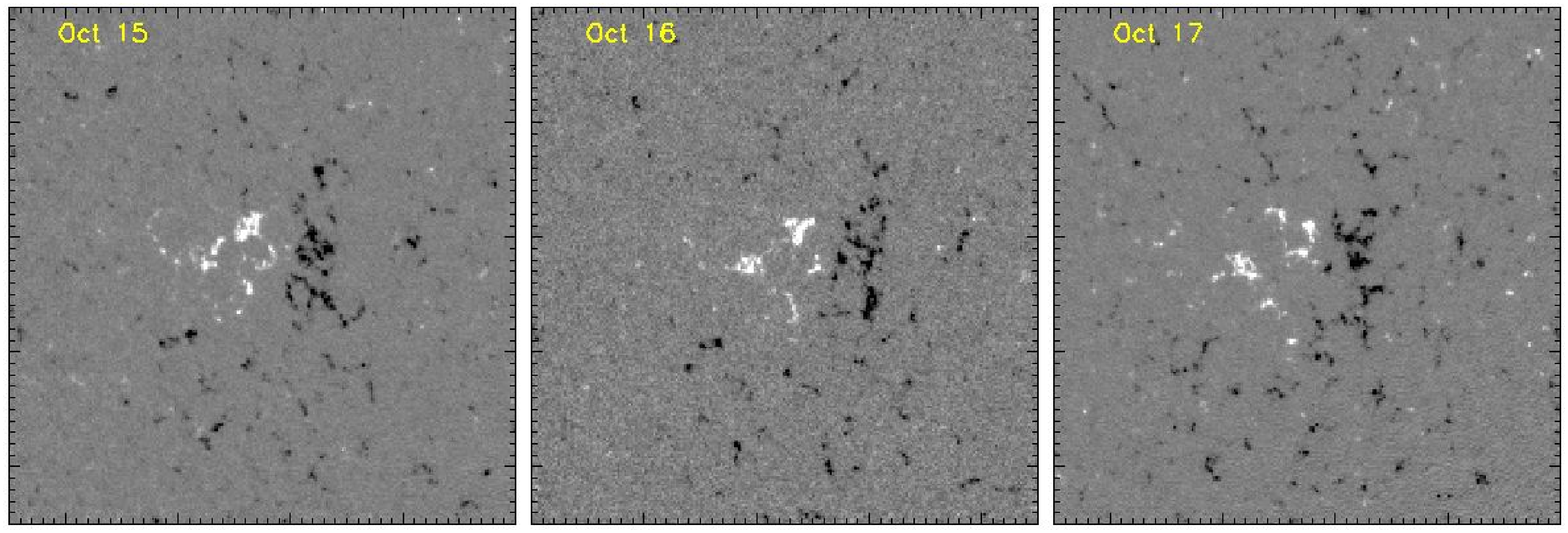}

\caption{Top panel:  SOHO/MDI magnetogram (left) and \emph{Hinode}/XRT Al$\_$mesh filter full-disk image (right) for 2007 October 17.  Bottom panel:  AR magnetic field evolution from October 15 to 17, 2007 (saturated at $\pm$250 G).  \label{context}}
\end{figure*}

A small `anemone' AR inside a low-latitude CH was observed close to central meridian on 2007 October 17.  Anemone ARs are usually associated with emerging flux within unipolar regions, especially CHs \citep[e.g.][]{asai08,baker09}. During emergence, the AR interacts with the ambient coronal field and the AR's magnetic connectivities are reorganized via interchange reconnection. Newly created magnetic loops extend radially from the location of the included AR polarity, thus creating the characteristic anemone configuration.  In this event, the AR's included positive polarity reconnects with the surrounding negative field of the CH, forming new compact loops on the AR's eastern side at the interface of the oppositely aligned field.  The new loops are evident in the \emph{Hinode}/XRT full-disk image in Figure \ref{context} (top right).  The overall magnetic configuration and 3-day on-disk temporal evolution of the AR--CH complex are shown in SOHO/MDI full-disk and zoomed magnetograms in Figure \ref{context}.  

\begin{figure*}
\epsscale{1.0}
\plotone{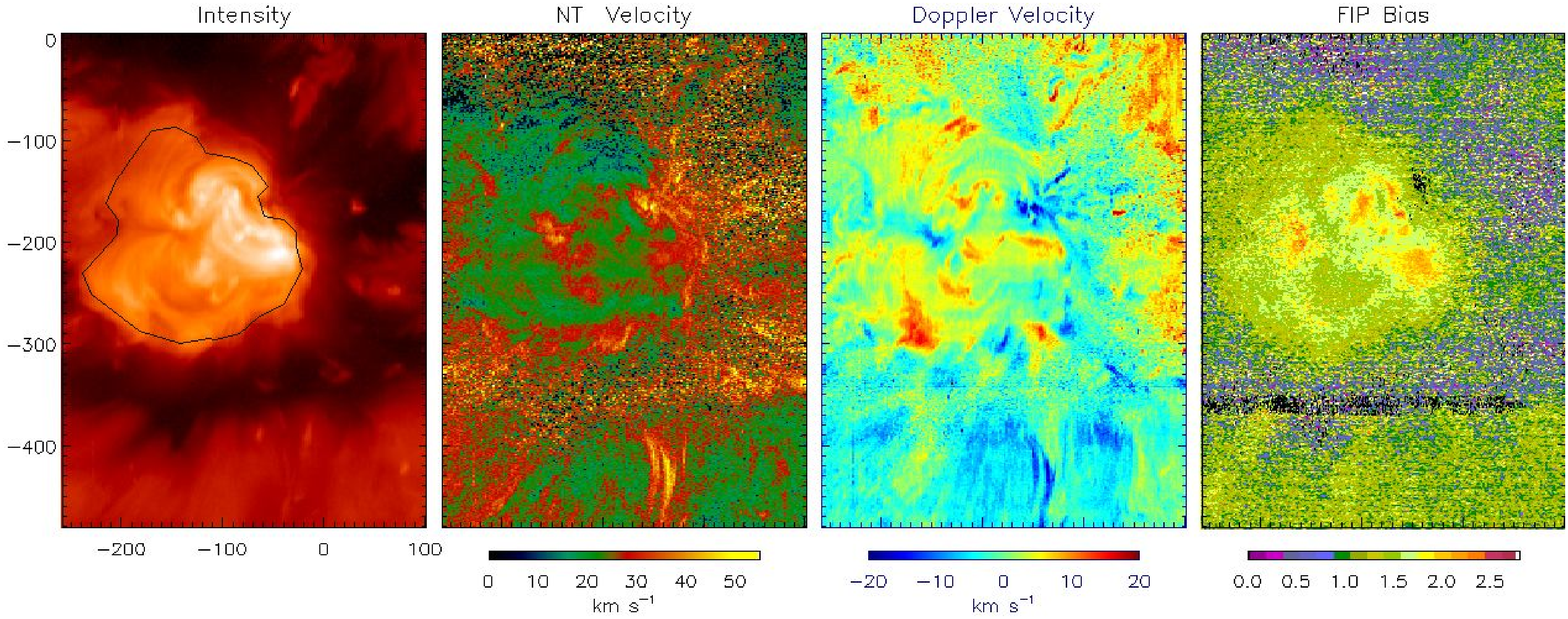}
\plotone{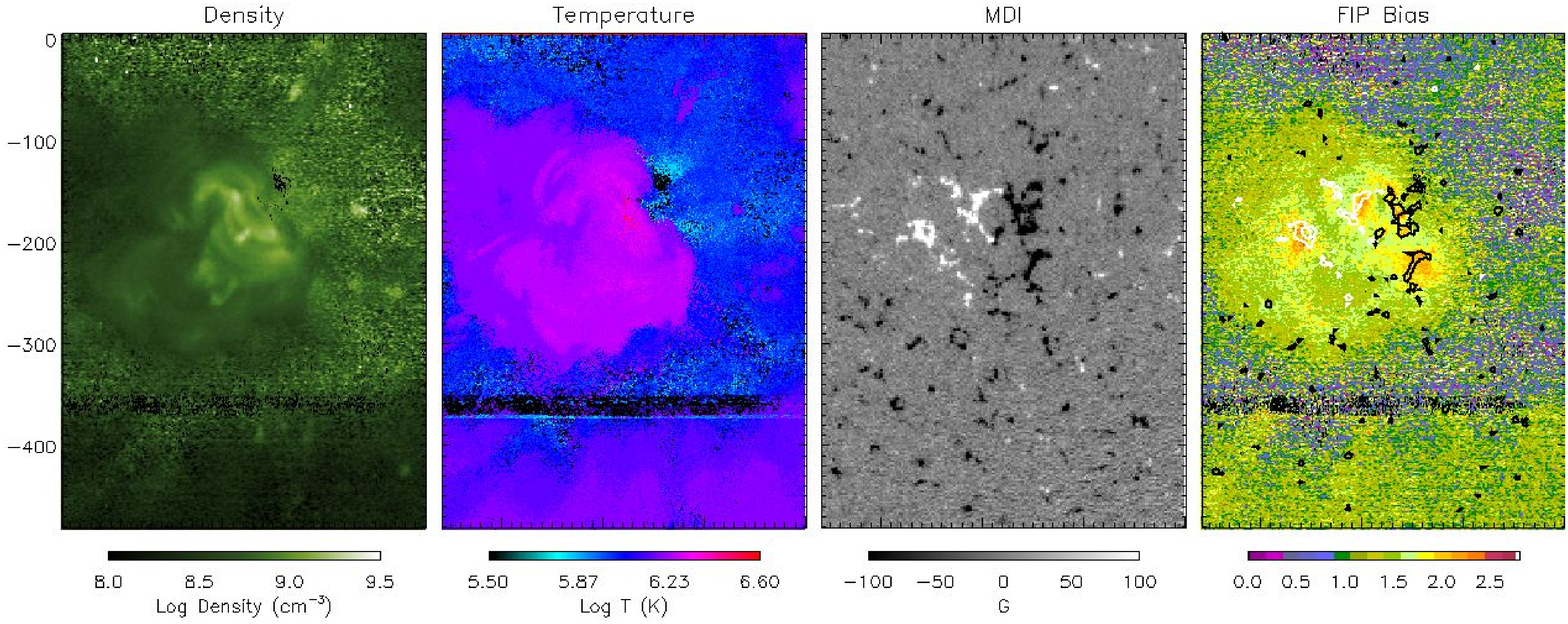}
\caption{Top panel (left to right):  EIS Fe {\sc xii} intensity, non-thermal velocity, Doppler velocity, and abundance maps for 2007 October 17 at 02:47 UT.  Bottom panel (left to right):  EIS Fe {\sc xiii} density and temperature maps, MDI magnetogram closest to the EIS raster time (saturation is $\pm$100 G), and abundance map overlaid with $\pm$100 G MDI contours.   X and Y axes are in arcsec. \label{all_eis}}
\end{figure*}

A complete description of this event is given in \cite{baker12}, however, in this study we focus on a single \emph{Hinode}/EIS raster scan timed at 02:47 UT on October 17.  EIS observed the AR--CH complex using the slit scanning mode with the 2$\arcsec$ slit and 2$\arcsec$ step size for 180 pointing positions to build up a FOV of 360$\arcsec$$\times$512$\arcsec$.  Total raster time of 2$\frac{1}{4}$ hours is comprised of 45 s exposure time at each pointing position.  Data reduction was carried out using standard SolarSoft EIS procedures. Raw data were corrected for dark current, hot, warm, and dusty pixels, and cosmic rays.  Instrumental effects of slit tilt, CCD detector offset, and orbital variation were corrected.  Calibrated spectra were fitted with a single Gaussian function.  Reference wavelengths were taken from the average value of a relatively quiescent region along the bottom of the raster.   Among the many emission lines simultaneously observed within the EIS spectral bands, we primarily use the Fe {\sc xii} 195.12 {\AA} line for intensity, Doppler and non-thermal velocity maps, Fe {\sc xiii} 202.02 {\AA} and 203.83 {\AA} line pair for the density map, and S {\sc x} 264.223 {\AA} and Si {\sc x} 258.375 {\AA} lines and various Fe ions for constructing the abundance  map.  EIS Fe {\sc xii} intensity, nonthermal and Doppler velocity, and S {\sc x}--Si {\sc x} abundance maps are shown in the top panel of Figure \ref{all_eis}.  Along the bottom panel of Figure \ref{all_eis}, the Fe {\sc xiii} density map, temperature map, MDI magnetogram, and abundance map overlaid with MDI $\pm$ 100 G contours are displayed.

\section{Abundance Maps}
\label{fip_map}

To construct the abundance map we first prepared coaligned intensity images for all the spectral lines we used. This was done by calculating the spatial displacement between short- and long-wavelength CCDs for each wavelength and extracting the common area (359$\arcsec$$\times$485$\arcsec$). We then fitted Gaussian functions to a series of  strong spectral lines from consecutive ionization stages of Fe {\sc viii-xvi}. Most of the lines used are unblended so were fit to single Gaussian functions. In a few cases the line is blended, e.g. Fe {\sc xii} 195.119 {\AA} \citep{delzanna&mason_2005}, so multiple Gaussian fits are more appropriate. We also fitted the S {\sc x} 264.223 {\AA} and Si {\sc x}  258.375 {\AA} lines to be used for the abundance measurement. 

The density in each pixel was then measured using the Fe {\sc xiii}  202.044/203.826 diagnostic ratio, and contribution functions for all the spectral lines were calculated assuming this density. We have used the CHIANTI database \citep{dere1997,landi12} for calculation of the contribution functions, adopting the photospheric abundances of \cite{grevesse07}. An emission measure (EM) distribution was then calculated for every pixel using a Markov-Chain Monte Carlo (MCMC) algorithm that calculates alternative solutions by randomly perturbing the observed intensities \citep{kashyap98}. The EM distributions are convolved with the contribution functions and fitted to the observed intensities to determine the best solution. We compute 100 realizations of the solution for each of the 170,000 pixels in the raster. 

The EM is computed using the ten low FIP Fe lines only. 
Since Si is also a low FIP element the derived EM should reproduce the intensity of the Si {\sc x}  258.375 {\AA} line well. Any mismatch, however, is removed by automatically scaling the EM to reproduce the Si {\sc x} line intensity. Once the best fit for each pixel is found, the FIP bias is calculated as the ratio of the predicted to observed intensity for the S {\sc x}  264.233 {\AA} line. Any temperature and density sensitivity of the ratio is accounted for by this method. 

\section{Results}
\label{results}
FIP bias of large-scale features within the AR--CH complex can be identified in the spatially resolved abundance map displayed in Figure \ref{all_eis}.  The surrounding CH is clearly photospheric in composition with a FIP bias of $\sim$1, in agreement with previous studies \citep{feldman98,brooks11}.  FIP bias levels within parts of the anemone AR are $>$2.  Though the overall FIP bias in this AR is somewhat low compared with previous studies of individual AR features,  (see Table \ref{fip_bias}), it is clearly above the CH FIP bias.  Quiet Sun FIP bias in the areas surrounding the outer edge of the CH varies between 1 and 1.5.  

Poor signal-to-noise prevents detailed analysis of FIP bias fine-structure within the CH, however, there is clear evidence of fine-structure within the anemone AR.  High FIP bias is concentrated in a few patches located very close to regions of relatively strong magnetic flux density at coronal loop footpoints.  This is demonstrated in Figure \ref{all_eis}, bottom right panel, where MDI contours of $\pm$100 G are overlaid on the abundance map.   FIP bias values are $\sim$2.5$-$3 in these regions.  In addition, slightly enhanced FIP bias, between 2 and 2.5, appears to trace coronal loops connecting opposite polarity magnetic flux.  

\begin{figure}
\epsscale{1.3}
\plotone{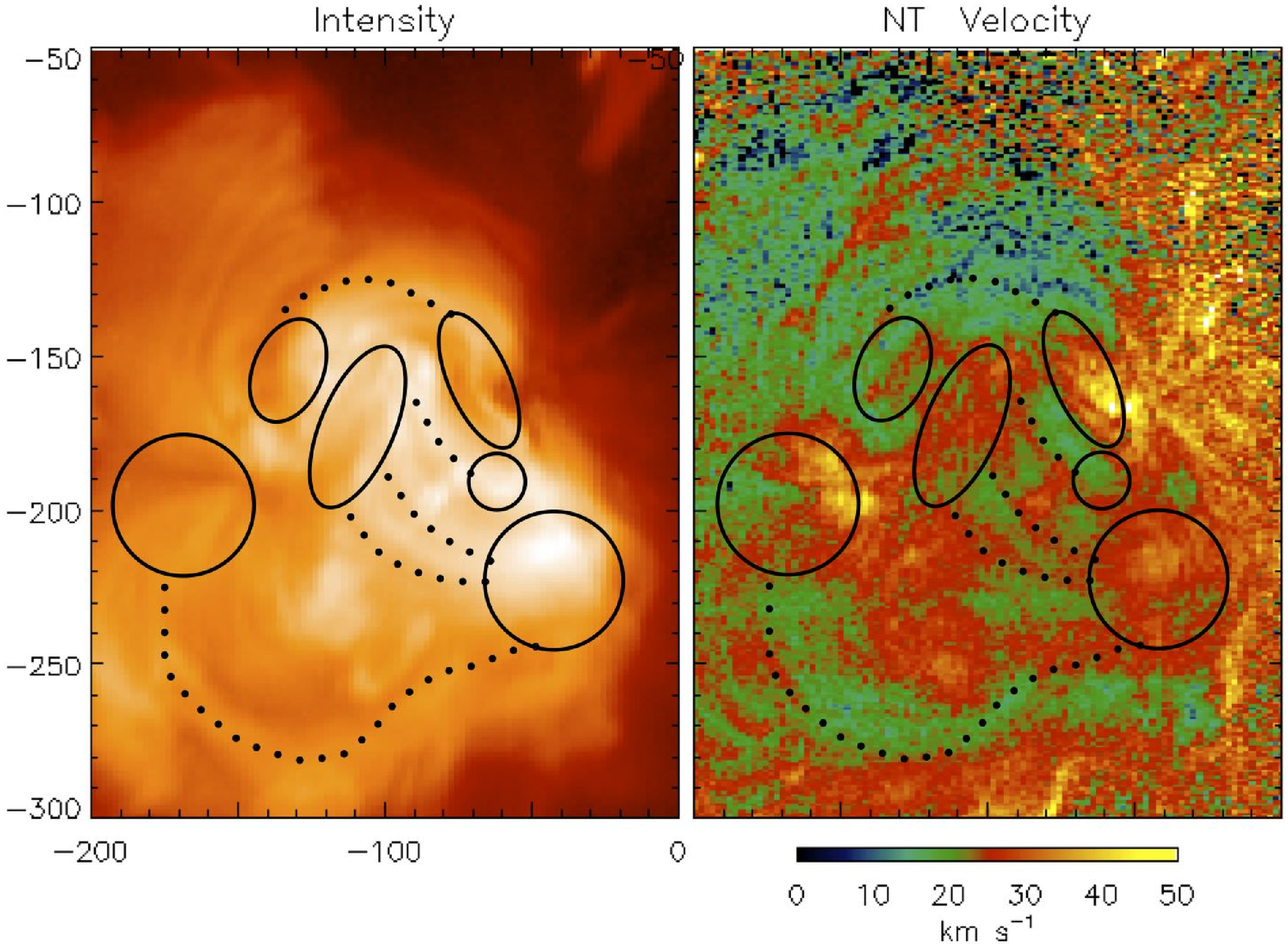}
\plotone{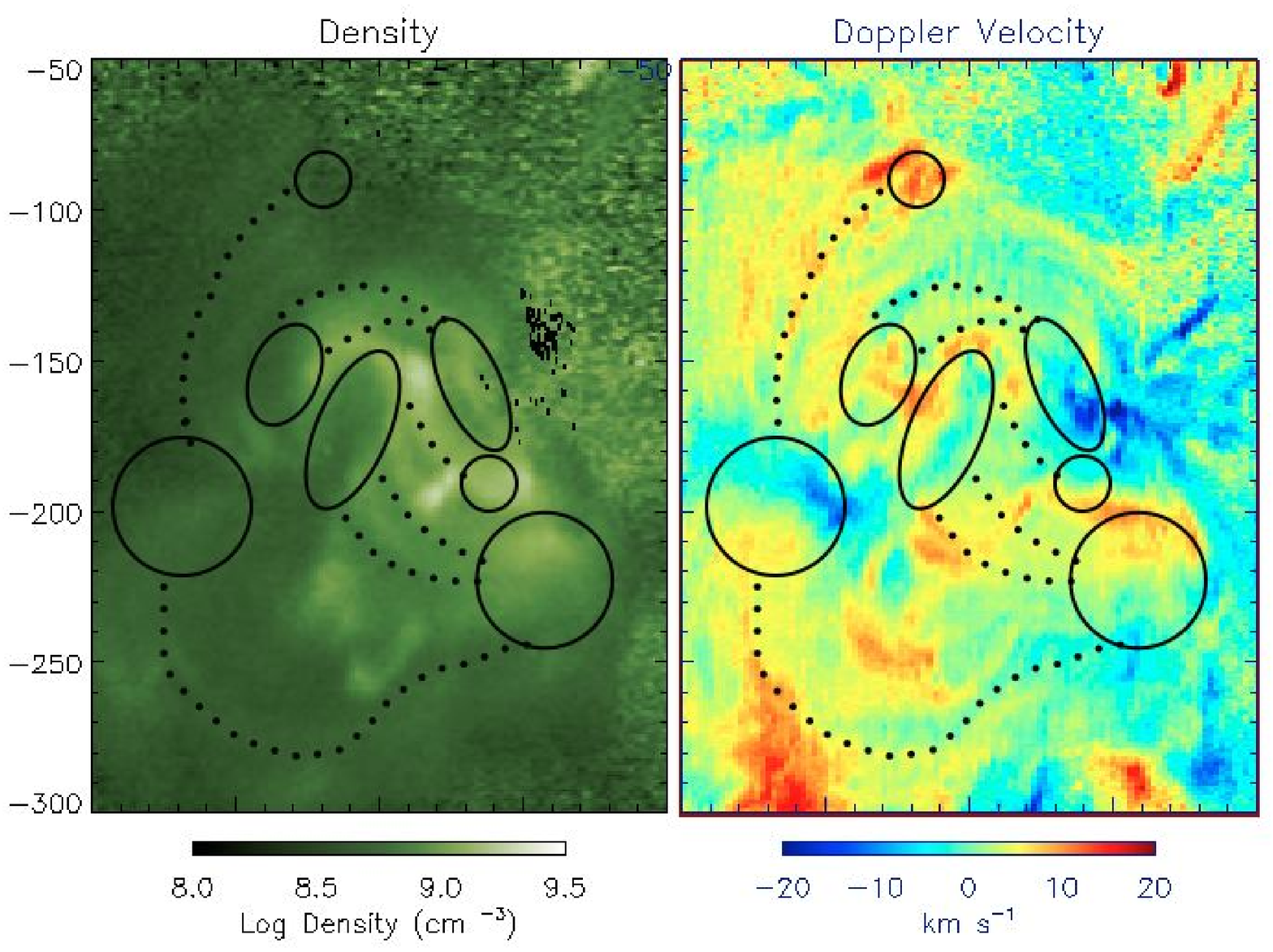}
\plotone{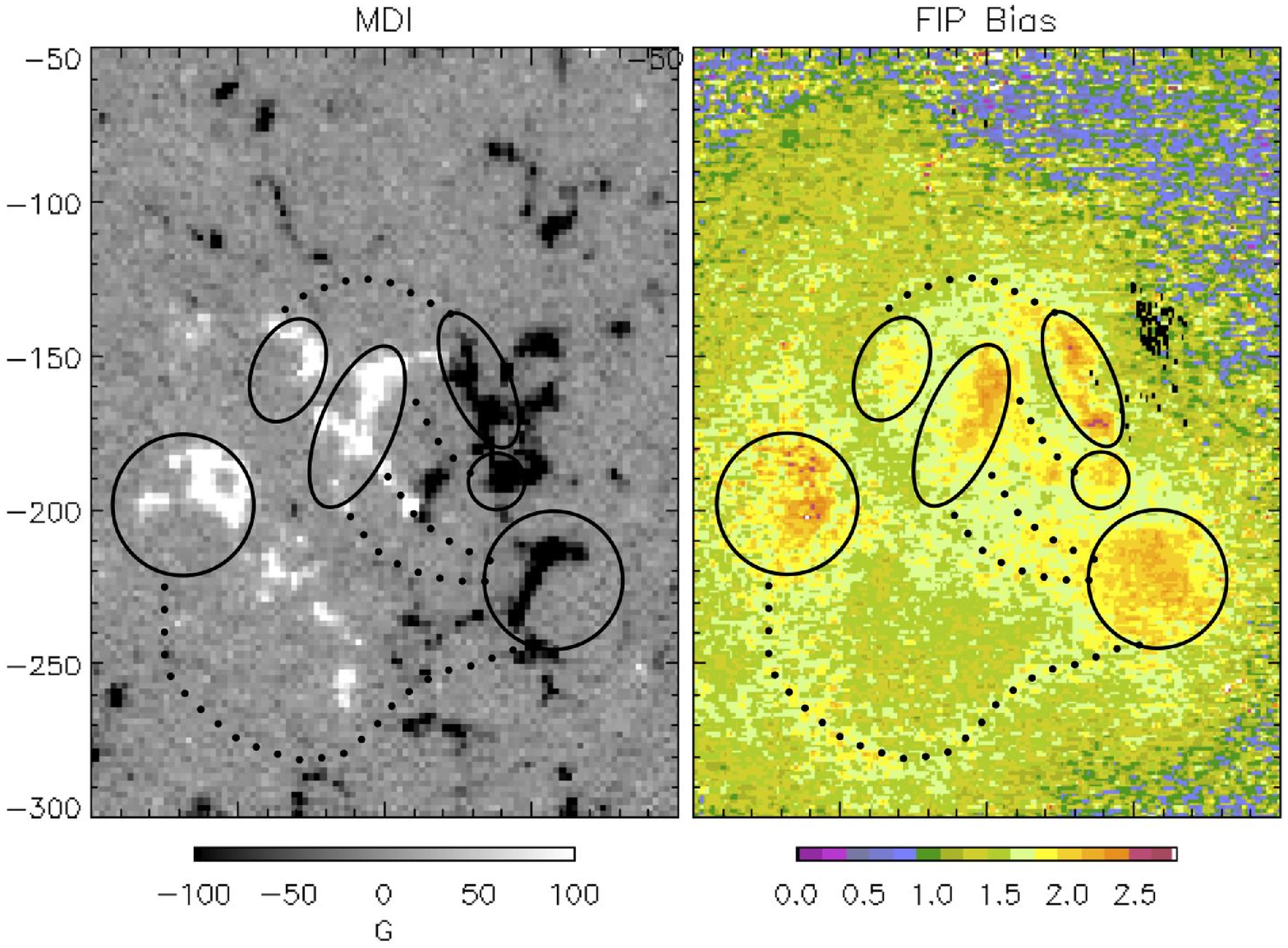}
\caption{Zoomed AR--CH complex Fe {\sc xii} intensity and nonthermal velocity maps (top panel), Fe {\sc xiii} density and Fe {\sc xii} Doppler velocity maps (middle panel), MDI magnetogram and abundance map (bottom panel), all showing locations of high FIP bias at loops footpoints (solid black ellipses) and loop traces of enhanced FIP bias (dotted black curves).   X and Y axes are in arcsec. \label{maps_over}}
\end{figure}

Comparison of the maps in Figure \ref{all_eis} suggests morphological groupings of FIP bias structures.  This is examined more closely in Figure \ref {maps_over}.  High FIP bias patches are spatially coincident with internal AR loop footpoints in the Fe {\sc xii} intensity and nonthermal maps.  The dotted lines of slightly enhanced FIP bias appear to trace very well loops connecting bipoles comprising the AR (see the animation abund$\_$intensity.mp4 included as an electronic supplement to this article).  These magnetic connections do not represent newly reconnected loops between emerging AR and surrounding CH loops but original AR connectivities.  They can be regarded as old loops.

\begin{figure}
\epsscale{0.7}
\plotone{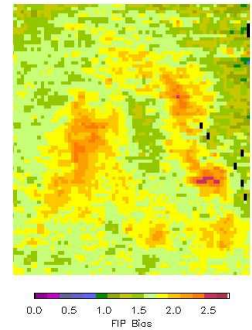}

\caption{Zoomed abundance map of the inverse S-shaped sigmoid channel of low FIP bias along the main PIL hosting a filament within the AR where a sigmoid/flux rope forms and eventually erupts \citep{baker12}. \label{sig}}
\end{figure}

The abundance map is further zoomed in Figure \ref{sig} to highlight a channel of low FIP bias in between the western-most ellipses to the AR's N/NW in the lower right panel of Figure \ref{maps_over}.  This region of lower FIP bias is cospatial with the main magnetic polarity inversion line (PIL) within the AR, along which the bright coronal loops become increasing sheared and sigmoidal in soft X-rays by 18:00 UT on the 17th and eventually erupt at 07:30 UT on the 18th \citep{baker12}. 

Possible links between FIP bias in the anemone AR and plasma parameters are better illustrated with correlation plots constructed using the maps in Figure \ref{all_eis}.  Data from pixels within a contour fitted around the AR were extracted for intensity, nonthermal and Doppler velocities, density, temperature and absolute magnetic flux density.  The contour is overplotted on the intensity map of the AR--CH complex in Figure \ref{all_eis}.  Pixels with a Doppler velocity in between  $\pm$5 km s$^{-1}$ were removed so that only relatively strong flows in excess of the EIS error are included.   Furthermore, a reduced chi-squared filter based on the EM calculation was applied to the data. The filter removes pixels that deviate too strongly from the fitted EM ($\bar{\chi}^2 < 2$) without being too stringent in the case of FIP bias values.  

\begin{figure*}
\epsscale{0.28}
\plotone{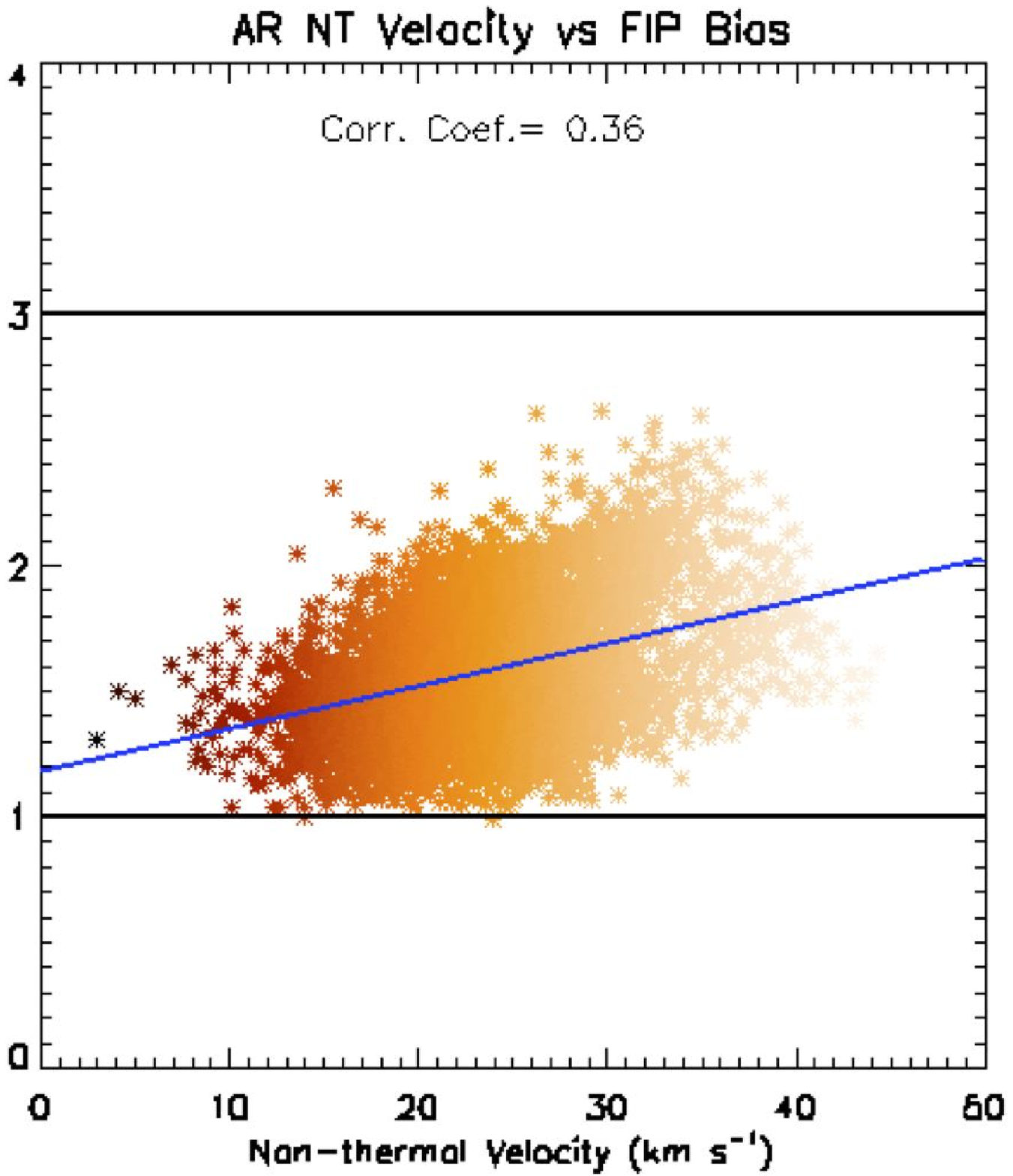}
\plotone{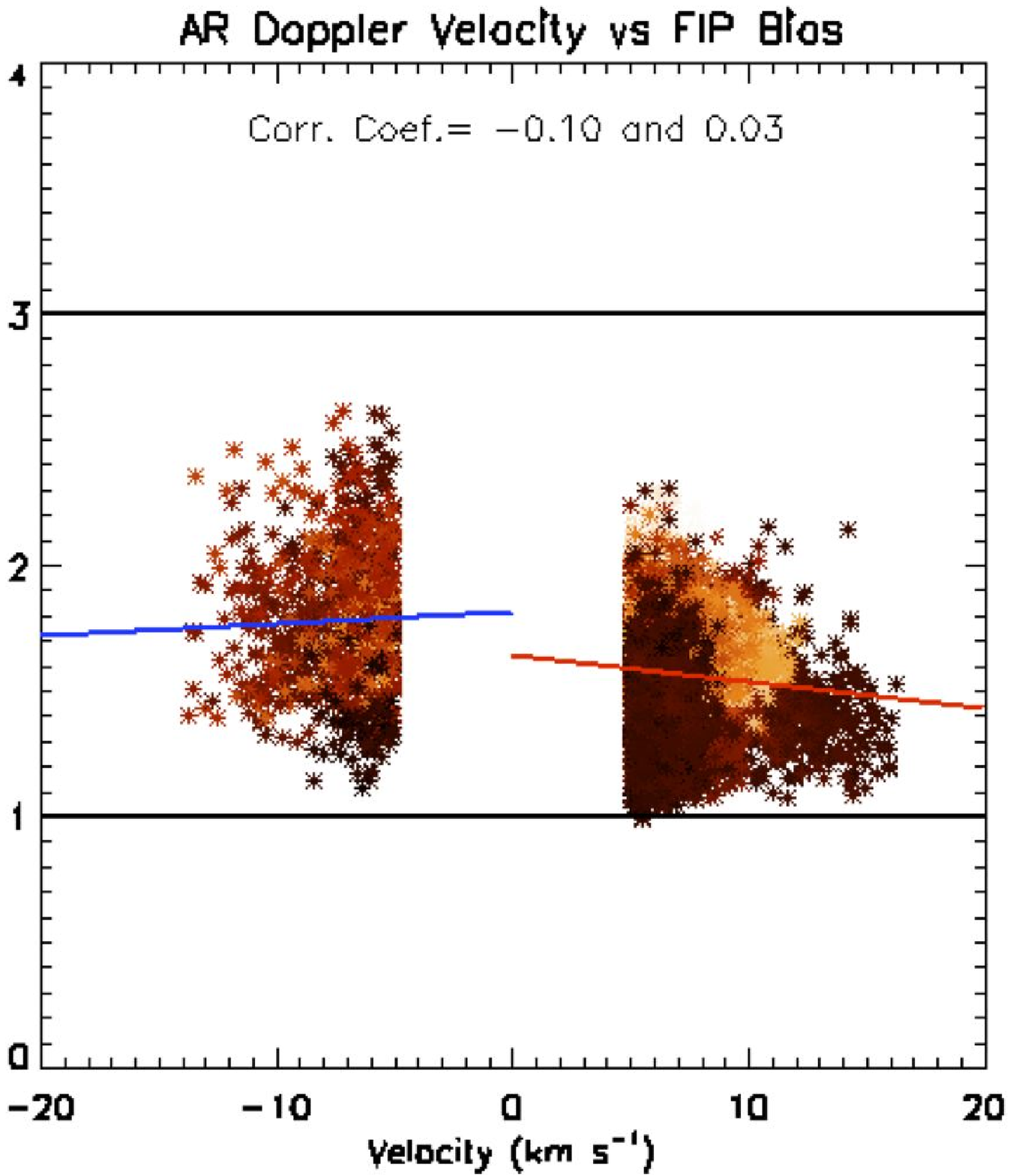}
\plotone{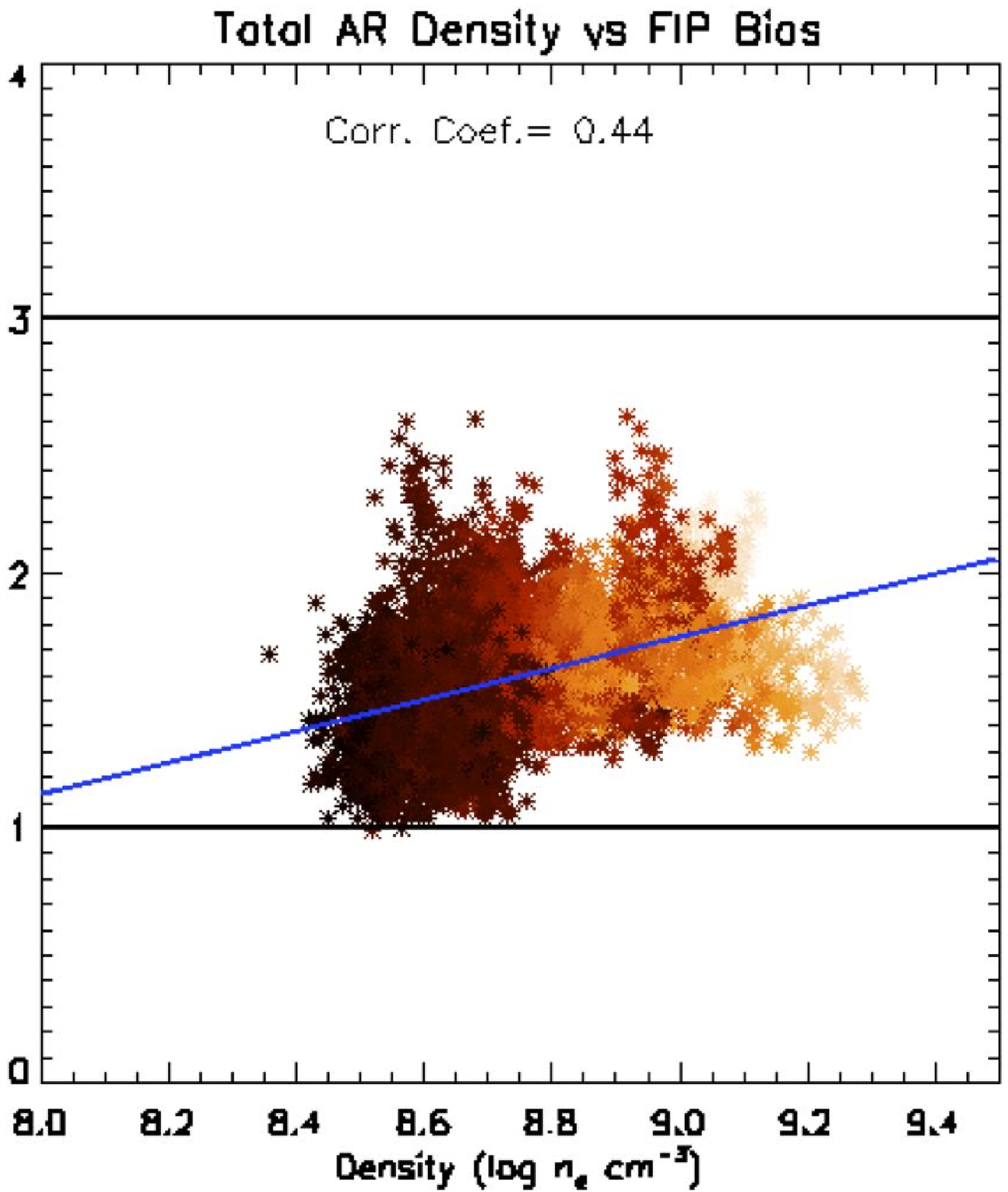}
\plotone{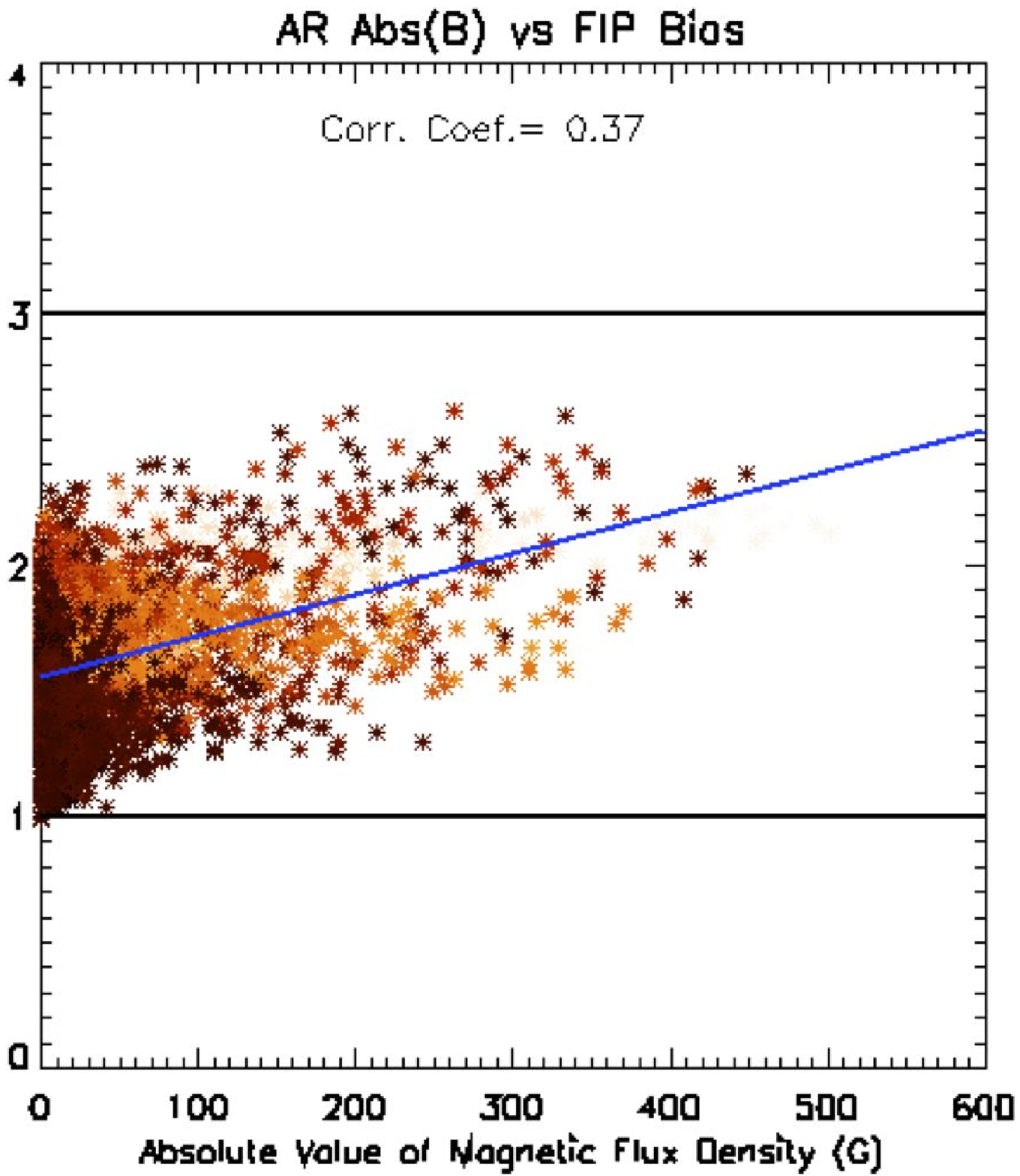}

\caption{Correlation plots of nonthermal velocity, Doppler velocity (upflows and downflows), plasma density, and absolute value of magnetic flux density vs. FIP bias.  Correlation coefficients are given in each plot.  Blue lines are linearly fitted lines to the data.  In Doppler velocity vs FIP bias plot, upflows/downflows have been fitted separately (blue/red lines).  The color scheme used for the data in each plot is based on the intensity level with the corresponding pixel of the intensity map in Figure \ref{all_eis}.  \label{corr_plots}}
\end{figure*}
 
Selected plasma parameter data are plotted vs. FIP bias in Figure \ref{corr_plots}.  The data are fitted with a linear relationship and correlation coefficients are given in each correlation plot.  Overall, there is a moderate positive correlation of three plasma parameters with FIP bias in the AR:  nonthermal velocity, (correlation coefficient of 0.36), density (0.44; intensity is similarly correlated with FIP bias), and absolute value of magnetic flux density (0.37).  Temperature is weakly correlated with FIP bias (correlation coefficient of 0.20; not shown).  Doppler velocity, whether downflow or upflow, does not appear to be correlated with FIP bias.  Note that the method utilized in constructing the abundance map is designed to remove any effect of temperature and density on FIP bias measurement; therefore, the moderate positive correlation of the measured FIP bias with density and the weak correlation with temperature are expected to be real and due to the underlying physics in the creation of FIP bias. 

\section{Discussion}
\label{discussion}

 Table \ref{fip_bias} shows typical FIP bias values for established ARs to be greater than four but for this event, the FIP bias enhancement is 2--3.  We suggest it is likely that the overall relatively low FIP bias levels measured within the AR are attributable to the age and magnetic configuration of the AR-CH complex.   In general, large variation in AR FIP bias is related to the variation in the average age of ARs \citep{widing01,mckenzie92}.  New flux emergence is characterized by photospheric composition \citep{sheeley95,sheeley96,widing97,widing01}.  Thereafter, the FIP bias of evolving ARs progresses at approximately a constant rate toward coronal levels within days of emergence \citep{widing01}.  For the studied AR, the low level of FIP bias may be due to the fact that the anemone AR is dominated by recently formed loops instead of the older, more extended, lower density loop structures of mature ARs \citep{young97}.  The age of the AR should not be much more than $\approx$ 10 days as its low flux content, 3$\times$10$^{21}$ Mx, puts it in the small AR category for which the average lifetime is measured in days \citep{schrijver08}.

The surrounding CH may also contribute to the low FIP bias levels in the dominantly negative monopolar field.  CH plasma undergoes very little, if any modification upon emerging from the photosphere therefore its composition remains unchanged at levels close to 1, as was confirmed by this study.  The fact that the AR is fully surrounded by CH plasma of photospheric composition means that low FIP bias plasma of the CH can readily mix with the high FIP bias of the AR plasma after interchange reconnection has occurred (forming part of the anemone AR loops).  Since anemone ARs form only in the unipolar field of CHs, it is plausible that they would have lower FIP bias levels compared with  ARs surrounded by mixed polarity QS plasma of higher FIP bias, assuming other factors such as age are similar between the ARs.  This is consistent with the results of \cite{wang09}, who found that over solar cycle 23, the average enrichment factor compared to photospheric values for a group of low FIP elements, including Si, was only 1.7--2.5 when the fraction of open flux on the visible side of the solar disk exceeded 14$\%$, however, the enrichment factor increased to 2.8--4.2 when the open flux fraction was less than 7$\%$.  

 The internal structure of the AR consists of higher FIP bias loops and a low FIP bias channel around the main PIL of the AR.  This low FIP bias region could apriori be due to local flux emergence all along the PIL, however, such emergence will only consist of small, local bipoles which would be covered over by a set of coronal loops.  The FIP bias in the overlying loops would be expected to be similar to the higher FIP bias levels found in other parts of the AR.  Moreover, emergence is not observed along the PIL \citep{baker12}, therefore, the emergence hypothesis cannot explain the observed low FIP bias.  
However, significant flux cancellation along the main AR PIL has been ongoing for several days prior to the time of EIS observation on the 17th. The flux cancellation creates a sigmoid as seen in soft X-rays and is observed at the photosphere as the disappearance of small magnetic bipoles at the PIL occurs  \citep{baker12}. This implies that reconnection is occurring in the lower atmosphere because very small loops are required for submergence along the PIL \citep{vanballegooijen89}.  This scenario has been used to show that sigmoid formation can indicate the build up of a flux rope in \cite{green11}. \cite{green09} have shown that the bottom of the flux rope in a sigmoidal region can be located low down in the solar atmosphere as expected from its formation via flux cancellation. This means that field lines running under the flux rope will have a bald patch (BP) topology, where field lines are tangent to the photosphere. Photospheric plasma is able to get access to BP field lines as a result of their creation via reconnection in the lower atmosphere.

Converging and/or shearing motions in the photosphere can then induce the formation of a current sheet all along the separatrix attached to the BP, and typically a current sheet is also formed above the BP \citep{low88}. Then, magnetic reconnection is theoretically expected, as was verified in MHD simulations \citep[e.g.][]{aulanier10}. Reconnection at the BP was found to be part of the flux rope build up process before the flux rope becomes unstable and a CME is launched. Magnetic reconnection at the BP has the particularity to occur in the cold part of the solar atmosphere and then to input photospheric plasma in upwardly curved field lines, so lifting up and heating photospheric material in the new-formed longer coronal loops. Such plasma is expected to have photospheric abundances. Indeed, in EUV brightenings, plasma along BP field lines has been shown to have photospheric FIP bias in contrast to quasi-separatrix layer (QSL) field lines \citep{fletcher01}.  The presence of a low FIP bias sigmoid channel in the anemone AR may indicate that reconnection low in the atmosphere leads to mixing of photospheric composition material with coronal plasma with higher FIP.  The photospheric FIP values previously determined from \emph{Skylab} observations for an eruptive filament (see Table \ref{fip_bias}) together with the typical presence of BPs below filaments \citep[e.g.][]{Aulanier98} are consistent with our results and supportive of a BP topology within the sigmoid channel.

Though it is possible that uncertainties in the method employed to determine FIP bias may be contributing to the low levels measured in the AR, the impact is likely to be minimal when weighed against real factors such as age and magnetic topology of the AR-CH complex.  We selected well tested lines and the atomic data for these lines are expected to be accurate to derive the FIP bias.  Also, EIS has the best temperature and density diagnostics of any coronal instrument to date.  

A further convincing test of the FIP bias results is the correspondences found with the observed coronal structures. First, the CH was determined to have photospheric composition.  Second, the largest enhanced FIP bias patches are located at the base of coronal loops.  Third, coronal loop connectivities are clearly traced in the abundance map.  Finally, the low FIP bias channel around the main AR PIL is consistent with the presence of a sigmoid with a BP topology.  All of these correspondences between FIP bias and coronal structures show that the FIP bias uncertainties are well below the observed range of FIP bias variations.

One caveat is that Si and S are close to the usually defined boundary between high and low FIP elements. Although the observed variations clearly show that the Si/S ratio is sensitive to FIP bias, some models indicate that Si fractionates relatively less than other low FIP elements and S fractionates relatively more than other high FIP elements. This suggests that the actual level of FIP bias could be underestimated \citep{laming12}. The observational picture is less clear, however. Evidence suggests that S, for example, behaves like a high FIP element in active regions \citep{lanzafame_etal2002}, but over-fractionates in quiet regions \citep{brooks_etal2009}. An anemone AR may represent some intermediate state between the two. Note that our method of scaling the Fe emission measure to that of Si allows a calibration of any under-fractionation of Si. In \citet{brooks11} we found that this scaling was less than 20\%, indicating a similar behavior for Fe and Si, and that the FIP bias values are accurate in many cases.

\section{Conclusions}
\label{conclusions}
In this paper we analyze an anemone AR inside an on-disk CH using observations from \emph{Hinode}/EIS, XRT and SOHO/MDI from 2007 October 17.  We constructed large-scale 359$\arcsec$$\times$485$\arcsec$ intensity, nonthermal and Doppler velocities, density, temperature and abundance maps of the AR-CH complex.   FIP bias in the surrounding CH is $\sim$1, consistent with previous composition studies of CHs, and FIP bias of the anemone AR is 2--3, which is lower than those values obtained from AR composition studies listed in Table \ref{fip_bias}.  

No consensus exists as to a single mechanism of cause for the FIP effect, however, it is generally accepted that fractionation takes place in the chromosphere where low FIP elements are mainly ionized and high FIP elements are at least partially neutral.  \cite{widing01} identified the footpoints and legs of loop-like structures as the location where fractionation and uplift occurs.  We find strong evidence in support of \cite{widing01} in the anemone AR, where high FIP bias is distinctly concentrated at the AR's loop footpoints (see ellipses in Figure \ref{maps_over}).  In a young AR, there is insufficient time for high FIP plasma to fill the coronal loops so the concentration of high FIP bias at the footpoints suggests that FIP bias enhancement begins at the AR's footpoints, in close proximity to where fractionation occurs.

We detect the start of high FIP bias plasma mixing in some of the coronal loops in the abundance map in Figure \ref{maps_over}.  Pathways of slightly enhanced FIP bias are traced along loops connecting opposite polarities of bipolar magnetic concentrations within the AR, indicating that the loops are partly filled with the high FIP bias plasma.  The degree of mixing of plasma along the loops is expected to be limited in this case due to the relatively weak heating generated by the weak mean magnetic field of the anemone AR which was measured to be $\approx$ 80 G when it crossed the solar central meridian \citep{baker12}.  Furthermore, when the anemone AR and nearby CH fields reconnect, new loops form which are similar in size to the AR loops so enhanced FIP bias transferred to the reconnected loops is partly mixed with previously open-field plasma of lower FIP bias.  The enhanced FIP bias is not entirely diluted because the loop size is similar to that of the pre-reconnection  loops.  


We also found a moderate positive correlation of high FIP bias with nonthermal velocity and the absolute value of magnetic flux density.  Our observations favor FIP effect models that are located in the vicinity of AR footpoints from where the fractionated plasma is then transported throughout the loops by diffusion.  One such model is the Laming FIP effect model \citep{laming04,laming09,laming12} which invokes the ponderomotive force arising from Alfv\'en waves of coronal origin reflecting from the chromosphere at loop footpoints and the induced generation of slow mode waves to explain FIP fractionation.  If we interpret the strong nonthermal velocity in the AR's footpoints to be a slow mode wave along the magnetic field then the slow mode amplitudes of the \cite{laming12} model are comparable to observed nonthermal velocities.  \citep[see] [Figure 8] {laming12}. 

 Finally, the low FIP bias observed within the inverse S-shaped sigmoid channel just above the AR's PIL is atypical of the global pattern of the AR FIP bias.  This low FIP bias cannot be due to emergence since it is not observed. Rather, we propose that low FIP bias in the sigmoidal channel could be the underbelly of a flux rope formed by flux cancellation along the PIL, which is highly suggestive of a BP topology.  In such a configuration, reconnection takes place at the photospheric level, lifting up photospheric plasma in the magnetic dips, implying a low FIP bias when mixed with the coronal plasma of the reconnecting loops.  To date, determining the particular magnetic topology of a sigmoid/flux rope configuration has proved to be problematic because both coronal loops and photospheric magnetic field are nearly aligned with the PIL so that it is difficult to determine if the magnetic configuration is normal or inverse (with curved down or up field lines, respectively).  FIP bias provides key evidence to clearly distinguish between BP and QSL topology, independent of direct observations of the  magnetic field.  This has far-reaching implications for predicting CMEs as BP reconnection is expected to be present well before a CME, as shown in MHD numerical simulations. Identifying a BP topology where reconnection is occurring allows for the  identification of the early building up of a flux rope which later will become unstable and will create a CME if the overlying magnetic field is not too strong (otherwise it is a failed eruption).   Then, we conclude that the abundance maps have the potential to identify the early formation of flux ropes which are potential sites of CMEs.  





\acknowledgements{Hinode is a Japanese mission developed and launched by ISAS/JAXA, collaborating with NAOJ as a domestic partner, NASA and STFC (UK) as international partners. Scientific operation of Hinode is by the Hinode science team organized at ISAS/JAXA. This team mainly consists of scientists from institutes in the partner countries. Support  for the post-launch operation is provided by JAXA and NAOJ (Japan), STFC (U.K.), NASA, ESA, and NSC (Norway).  LvDG and KS acknowledge the European Community FP7/2007-2013 programme through the eHEROES Network (EU FP7 Space Science Project No. 284461). KS also acknowledges support from the SWIFF Network (EU FP7 Space Science Project No. 263340).   LvDG acknowledges the Hungarian government for grant OTKA K-081421.  The work of DHB was performed under contract with the Naval Research Laboratory and was funded by the NASA Hinode program.  JC thanks UCL and the Max Planck Institute for an Impact Studentship award.  DB's work was supported by STFC.  We thank the referee for his/her constructive and inspiring comments which have improved the article.}

\bibliographystyle{apj}
\bibliography{test} 



\end{document}